\documentclass[12pt,twoside,pointlessnumbers,nofootinbib,smallheadings]{revtex4}

\usepackage[CJKbookmarks]{hyperref}
\usepackage{CJK}
\usepackage{amsmath}
\usepackage{graphics}
\usepackage{slashed}
\usepackage{graphicx}
\usepackage{indentfirst}
\usepackage{amssymb}
\usepackage{cancel}
\usepackage{leftidx}
\usepackage{simplewick}
\usepackage[page,title,titletoc,header]{appendix}

\begin{document}

\begin{CJK*}{GBK}{song}

\title{Testing CP-Violation in the Scalar Sector at Future $e^+e^-$ Colliders}

\author{Gang Li $^{2,}$\footnote{gangli@pku.edu.cn}, Ying-nan Mao $^{1,}$\footnote{maoyn@ihep.ac.cn},
Chen Zhang $^{2,}$\footnote{larry@pku.edu.cn}, and Shou-hua Zhu $^{2,3,4,}$\footnote{shzhu@pku.edu.cn}}

\affiliation{
$ ^1$ Center for Future High Energy Physics $\&$ Theoretical Physics Division,
Institute of High Energy Physics, Chinese Academy of Sciences, Beijing 100049, China \\
$ ^2$ Institute of Theoretical Physics $\&$ State Key Laboratory of
Nuclear Physics and Technology, Peking University, Beijing 100871,
China \\
$ ^3$ Collaborative Innovation Center of Quantum Matter, Beijing 100871, China \\
$ ^4$ Center for High Energy Physics, Peking University,
Beijing 100871, China
}

\begin{abstract}

We propose a {\em model-independent} method to  test CP-violation in the scalar sector through measuring the inclusive
cross sections of $e^+e^-\rightarrow Zh_1,Zh_2,h_1h_2$ processes with the recoil mass technique, where $h_1, h_2$ stand for
the 125 GeV standard model (SM) like Higgs boson and a new lighter scalar respectively. This method effectively measures
a quantity $K$ proportional to the product of the three couplings of $h_1ZZ,h_2ZZ,h_1h_2Z$ vertices.
The value of $K$ encodes a part of information about CP-violation in the scalar sector.
We simulate the signal and backgrounds for the processes mentioned above with
$m_{2}=40\textrm{GeV}$ at the Circular Electron-Positron Collider (CEPC) with the integrated luminosity $5\textrm{ab}^{-1}$.
We find that the discovery of both $Zh_2$ and $h_1h_2$ processes at $5\sigma$ level will indicate an $\mathcal{O}(10^{-2})$ $K$ value
which can be measured to $16\%$ precision. The method is applied to the
weakly-coupled Lee model in which CP-violation can be tested either before or after utilizing a ``$p_T$ balance" cut (see \autoref{simu}
for the definition). Lastly we point out that $K\neq 0$ is a sufficient but not a necessary
condition for the existence of CP-violation in the scalar sector, namely $K = 0$ does not imply CP conservation in the scalar sector.

\end{abstract}

\date{\today}

\maketitle

\newpage

\section{Introduction}
CP-violation was first observed through $K^0_L\rightarrow\pi\pi$ decay in 1964 \cite{CPVdisc}. More CP-violation effects
have been discovered in K- and B- meson sectors since then \cite{PDG}. In 1973, Kobayashi and Maskawa propose \cite{KM} that if there exist three
or more generations of fermions, one or more nontrivial phase(s) will be left in the quark mixing matrix, namely the
Cabibbo-Kobayashi-Maskawa (CKM) matrix \cite{KM,CKM}. In the standard model (SM), only a single nontrivial phase is left
which turns out to explain all the measured CP-violation effects successfully \cite{PDG}. However, it is still
necessary and attractive to study additional sources of CP-violation, which may help to
understand the matter-antimatter asymmetry in the universe \cite{PDG,Plank}.

In the SM, there is no CP-violation in the scalar sector. In models
with additional scalars, extra CP-violation may be introduced in the scalar sector \cite{2HDM}. For example, in a minimal
extension of SM \cite{singlet}, some kinds of two-Higgs-doublet models (2HDM) like Lee model \cite{Lee} or Georgi model \cite{georgi},
and Weinberg model which contains three Higgs doublets \cite{3HDM}, etc., there exists CP-violation in the scalar sector.
In such models, a Higgs boson can be a CP-mixing state. As an example, two of the authors have studied the phenomenology of
Lee model which contains spontaneous CP-violation in the scalar sector in detail \cite{our1,our2,thesis}. These papers revealed
the possible correlation between the lightness of Higgs boson and the smallness of CP-violation based on spontaneous CP-violation
mechanism which provides another important motivation to study CP-violation further in the scalar sector.

In 2012, a SM-like Higgs boson was discovered by the ATLAS and CMS collaborations \cite{CMS,ATLAS} with its
mass around 125 GeV \cite{higgsmass}. Its spin and CP properties have also been studied through the final state distributions of
$h\rightarrow ZZ^*\rightarrow4\ell$ decay process with the conclusion that a pure $0^+$ state is favored and
a pure $0^-$ state is excluded at over $3\sigma$ level \cite{spinCP1,spinCP2,spinCP3}. However, a CP-mixing state is
still allowed \cite{spinCP1,spinCP4} because the contribution from pseudoscalar component is loop induced and thus
highly suppressed.

CP-violation beyond the SM may show several kinds of indirect effects \footnote{Here ``indirect" means these phenomena will
show evidence for CP-violation, but we cannot extract the CP-violation vertex through these processes; while in the ``direct"
effects discussed below, we can obtain the CP-violation vertex through these measurements directly. Besides the effects discussed
below, the Higgs cubic self coupling could also be modified \cite{hsc} though the modification does not imply CP-violation.}.
For example, it may contribute to the
electric dipole moments (EDM) of election or neutron \cite{EDM} which are stringently constrained experimentally
\cite{eEDM,nEDM}; it may contribute to meson mixing matrix element and thus a modification from SM prediction could occur
\cite{mixing}; or it may also contribute to the anomalous $ZZZ$ coupling vertex \cite{GH,ZZZ1} which could lead to a nontrivial
CP-sensitive asymmetry in $e^+e^-\rightarrow ZZ$ process \cite{ZZZ2}.

However, to study the exact sources of extra CP-violation, we need their direct effects. For example, a CP-mixing Higgs boson
could couple to a fermion through the effective interaction
\begin{equation}
\mathcal{L}_{hf\bar{f}}=-h\bar{f}(g_S+\textrm{i}g_P\gamma^5)f,
\end{equation}
where $g_S$ and $g_P$ may be of the same order. For $f=\tau$, it is possible to test CP-violation effects in $h\tau^+\tau^-$
vertex at future $pp$ or $e^+e^-$ colliders \cite{CPtau1,CPtau2,CPtau3} using the final state distribution of $h\rightarrow\tau^+\tau^-
\rightarrow\nu\bar{\nu}+X$ decay process. Similarly, for $f=t$, the top polarization asymmetry in $e^+e^-\rightarrow t\bar{t}h$
process is useful to test CP-violation effects in $ht\bar{t}$ vertex \cite{CPt}.

In this paper, we will focus on the scalar sector itself and propose a {\em model-independent} method to test CP-violation effects
in the scalar sector through the interaction between scalars and massive gauge bosons. The paper is organized as follows. In \autoref{method}
we describe our method and perform a simulation study at the CEPC. In
\autoref{leemodel} we apply this method to the weakly-coupled Lee model.
And in \autoref{conc} we give our conclusions and discussions.

\section{Model-Independent Method to Test CP-Violation in the Scalar Sector at Future $e^+e^-$ Colliders}
\label{method}
If more than one neutral scalars are discovered in the future, the tree level interaction between neutral scalars and
massive gauge bosons could be written as
\begin{equation}
\label{hV}
\mathcal{L}_{\textrm{tree}}=\mathop{\sum}_ic_ih_iv\left(\frac{g^2}{2}W^{+\mu}W^-_{\mu}+\frac{g^2}{4c_W^2}Z^{\mu}Z_{\mu}\right)
+\mathop{\sum}_{i<j}\frac{c_{ij}g}{2c_W}Z_{\mu}\left(h_i\partial^{\mu}h_j-h_j\partial^{\mu}h_i\right).
\end{equation}
Here $g$ is the $\textrm{SU}(2)_L$ coupling constant, $c_W$ denotes the cosine of electro-weak angle $\theta_W$ \footnote{In this paper,
we denote $s_{\alpha}\equiv\sin\alpha$, $c_{\alpha}\equiv\cos\alpha$, and $t_{\alpha}\equiv\tan\alpha$ for any angle $\alpha$.},
$v$ is the vacuum expected value for
SM scalar field, and $h_i$ represents the $i$th scalar. For the first two terms, a nonzero tree-level $h_iVV$ vertex requires
that $h_i$ must contain CP-even component; while for the last term, a nonzero tree-level $h_ih_jZ$ vertex requires that
$h_i$ and $h_j$ must contain components with different CP-properties. If CP is a good symmetry, there must be some
terms vanishing in (\ref{hV}); on the other hand, if all $c_i$ and $c_{ij}$ are nonzero, there must be CP-violation in
the scalar sector.

\subsection{Method for the Minimal Case}
For the minimal case, two neutral scalars with non-degenerate masses are required to be discovered. CP-violation can be confirmed with $c_1$, $c_2$, and
$c_{12}$ all measured to be nonzero. It is natural to define
\begin{equation}
\label{kdef}
K\equiv c_1c_2c_{12}
\end{equation}
which is a useful quantity to measure the CP-violation effect
since $K\neq0$ is a sufficient condition for the existence of CP-violation in the scalar sector \footnote{One should  aware
that $K\neq0$ is not a necessary condition for the existence of CP-violation in the scalar sector which means in some models,
there may be CP-violation in the scalar sector with $K=0$, see the discussions in the last section.}.
As an example, in 2HDMs, there are three neutral Higgs bosons. We can use this idea to search for direct CP-violation effect once two
of them are discovered. A straightforward calculation shows $c_{12}=c_3$, and $K$ is just the product for all $c_i$ in 2HDM. That is
an important quantity to measure CP-violation in the scalar sector \cite{GH,ZZZ1,ZZZ2,K}.

At the LHC, the 125 GeV Higgs boson $h_1$ has already been discovered and the direct $h_1VV$ vertices have been confirmed
\cite{spinCP1,hVV}. If another Higgs boson $h_2$ is discovered and it has tree level
\footnote{In some special models, for example, the loop-philic model \cite{loop}, a loop-induced decay channel can also have a
large branching ratio even it is weakly-coupled.} decay channels $h_2\rightarrow WW,ZZ,Zh_1$,
it would strongly suggest CP-violation in the scalar sector which has already been discussed in \cite{our1,thesis,softCP}.
However, the $\sigma\cdot\textrm{Br}$ measurements at LHC depend on not only $c_{1,2}$ and $c_{12}$, but also a lot of other
parameters which would affect on the production cross section or branching ratios. Thus it is difficult to extract or constrain
the value of $K$ from these measurements without model-dependent assumptions.

\begin{figure}[h]
\caption{Feynman diagrams for associated production processes $e^+e^-\rightarrow Zh_1,Zh_2,h_1h_2$.}\label{FD}
\includegraphics[scale=1]{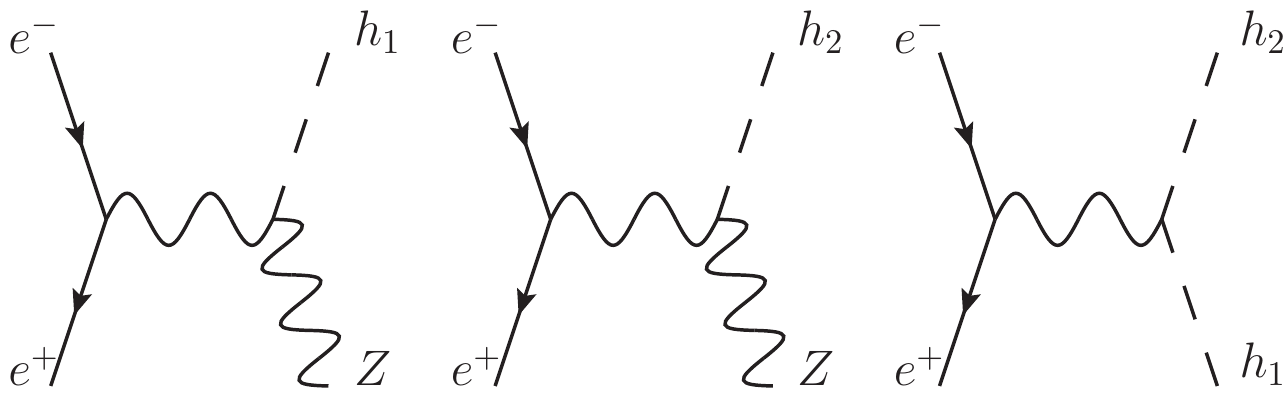}
\end{figure}

At future $e^+e^-$ colliders, we can use three associated production processes, $e^+e^-\rightarrow Z^*\rightarrow Zh_1,Zh_2,h_1h_2$,
to search for CP-violation in the scalar sector. The Feynman diagrams are shown in \autoref{FD}.
The cross sections at tree-level are given as \cite{csx1,csx2}
\begin{eqnarray}
\sigma_{Zh_i}&=&\frac{\pi\alpha^2s\cdot c^2_i}{96(s-m^2_Z)^2}\left(\frac{8s^4_W-4s^2_W+1}{s^4_Wc^4_W}\right)
\left(f^3\left(\frac{m^2_i}{s},\frac{m^2_Z}{s}\right)+\frac{12m^2_Z}{s}f\left(\frac{m^2_i}{s},\frac{m^2_Z}{s}\right)\right);\\
\sigma_{h_ih_j}&=&\frac{\pi\alpha^2s\cdot c^2_{ij}}{96(s-m^2_Z)^2}\left(\frac{8s^4_W-4s^2_W+1}{s^4_Wc^4_W}\right)
f^3\left(\frac{m^2_i}{s},\frac{m^2_j}{s}\right).
\end{eqnarray}
Here $s$ is the square of total energy in the center-of-mass frame, $s(c)_W$ denotes the (co)sine of electro-weak angle $\theta_W$,
and the function
\begin{equation}
f(x,y)\equiv\sqrt{1+x^2+y^2-2x-2y-2xy}.
\end{equation}
The cross sections are sensitive to $c_i$ or $c_{ij}$, but besides these, they don't depend on more details of the model.

The recoil mass technique \cite{rec1,rec2,CEPC} would be very effective for precision measurements on these
inclusive cross sections. For $e^+e^-\rightarrow Z(f\bar{f})h_i$ process, the
recoil mass is defined as \cite{rec1,rec2}
\begin{equation}
\label{rec}
m_{\textrm{rec}}\equiv\sqrt{s+m^2_{f\bar{f}}-2\sqrt{s}(E_f+E_{\bar{f}})}
\end{equation}
whose distribution would show a narrow peak around $m_i$ where $m^2_{f\bar{f}}=m^2_Z$ is the invariant mass
of the fermion pair. With this method, the sensitivity to $Zh_1$ inclusive cross section would
reach better than $1\%$ at future Higgs factories \cite{CEPC,TLEP,rec3} with $\sqrt{s}=250\textrm{GeV}$ and
$\mathcal{O}(\textrm{ab}^{-1})$ luminosity. The result doesn't depend on the decay channels of Higgs boson
which means this is a model-independent technique to measure $h_iZZ$ couplings $c_i$. Generalizing this technique
to $e^+e^-\rightarrow h_1(b\bar{b})h_2$ process, with $h_1$ the 125 GeV Higgs boson and $m^2_{b\bar{b}}=m^2_1$,
the distribution of $m_{\textrm{rec}}$ would show a narrow peak around $m_2$ and thus we can measure the
$e^+e^-\rightarrow h_1h_2$ inclusive cross section to extract the $h_1h_2Z$ coupling $c_{12}$ in a model-independent way
\footnote{In order to measure $\sigma_{h_1h_2}$ using this method, $\textrm{Br}(h_1\rightarrow b\bar{b})$ is needed as a
model-dependent quantity, which can be accurately measured
through $e^+e^-\rightarrow Zh_1$ process.}. Thus through measuring the three inclusive associated production cross
sections, we can extract all the three couplings $c_1,c_2,c_{12}$ and subsequently obtain $K$ in a model-independent way.

\subsection{Model-Independent Simulation Study}
\label{simu}
Here we perform a simulation study of the signal and backgrounds for the case $m_2=40\textrm{GeV}$ at
Circular Electron-Positron Collider (CEPC) \cite{CEPC} which would be a
$e^+e^-$ collider with $\sqrt{s}=250\textrm{GeV}$ \footnote{If the extra scalar is a heavier one, we can utilize this method
at $e^+e^-$ colliders with larger $\sqrt{s}$, like the International Linear Collier (ILC) \cite{ILC}.}.
Such a light scalar can occur in many models, such as 2HDMs \cite{our2,thesis,2HDM,2HDM1,2HDM2}.

Assuming $h_1$ is SM-like, $c_1\sim1$ which is consistent with the recent 125 GeV Higgs measurements \cite{recent}.
In the following we focus on the inclusive measurements on $Zh_2$
and $h_1h_2$ associated production processes. The strictest direct constraints on $c_2$ and $c_{12}$ came from
LEP results \cite{LEP1,LEP2} which give
\begin{equation}
|c_2|<0.18,\quad\quad|c_{12}|<0.54
\end{equation}
for $m_2=40\textrm{GeV}$ at $95\%$ C.L. assuming all scalars decay only to $b\bar{b}$ final states.

In our simulation analysis, we use WHIZARD-2.3.1 \cite{whi} to generate signal and background events with
initial state radiation (ISR) and beamstrahlung effects. For beamstrahlung effects, we use the built-in spectra CIRCE2
for the CEPC project \cite{CIRCE2}. For both processes, we adopt the recoil mass method in which we do not reconstruct
$h_2$ directly using its decay final states thus the results do not depend on the properties of $h_2$ except its mass.

For $Zh_2$ process, we choose the $Z\rightarrow\mu^+\mu^-$ decay channel. The corresponding backgrounds are $e^+e^-\rightarrow\mu^+\mu^-X$
where $X=e^+e^-$, $\mu^+\mu^-$, $\tau^+\tau^-$, $q\bar{q}$, $b\bar{b}$, $\nu\bar{\nu}$, or $\gamma\gamma$ \cite{CEPC,BKG1,BKG2,BKG3}.
We impose the basic cuts as \cite{CEPC,BKG1}
\begin{eqnarray}
&|\cos\theta_{\mu^{\pm}}|<0.98,\quad m_{\mu^+\mu^-}>15\textrm{GeV},\quad m_{\textrm{rec}}>15\textrm{GeV},\nonumber\\
&|\cos\theta_{e^{\pm},\gamma}|<0.995,\quad E_{\gamma}>0.1\textrm{GeV},\quad \Delta R_{ij}>0.4.
\end{eqnarray}
where $m_{\textrm{rec}}$ is defined in (\ref{rec}) with $f=\mu$ and $\Delta R_{ij}\equiv\sqrt{(\eta_i-\eta_j)^2+(\phi_i-\phi_j)^2}$
with $i$ and $j$ running over all partons in the final state \footnote{The cuts in the second line are useful to avoid the
infrared and collinear divergences in background processes. We do not consider the decays of $\tau$ leptons in our analysis. The final state with single photon can be totally rejected by the requirement of a large recoil mass $m_{\textrm{rec}}$ at the parton level.}. The transverse momentum of muon
is smeared by a Gaussian distribution with the standard deviation of \cite{CEPC}
\begin{equation}
\sigma_{1/p_T}=2\times10^{-5}\oplus1\times10^{-3}/(p_T\sin\theta)[\textrm{GeV}^{-1}].
\end{equation}

\begin{figure}[h]
\caption{Normalized kinematical distributions of the signal and backgrounds in the $e^+e^-\rightarrow Zh_2$ channel after the basic
cuts are applied. The first three figures show the $\cos\theta_{\mu^-}$, $p_T(\mu^+\mu^-)$, and $m_{\textrm{rec}}$ distributions
respectively in which we reconstructed only $\mu^+$ and $\mu^-$. The last figure shows the ``$p_T$ balance" distribution (see the
text below for details) in which we must tag at least one photon that breaks the inclusiveness a little bit.}\label{Zh2}
\includegraphics[scale=.182]{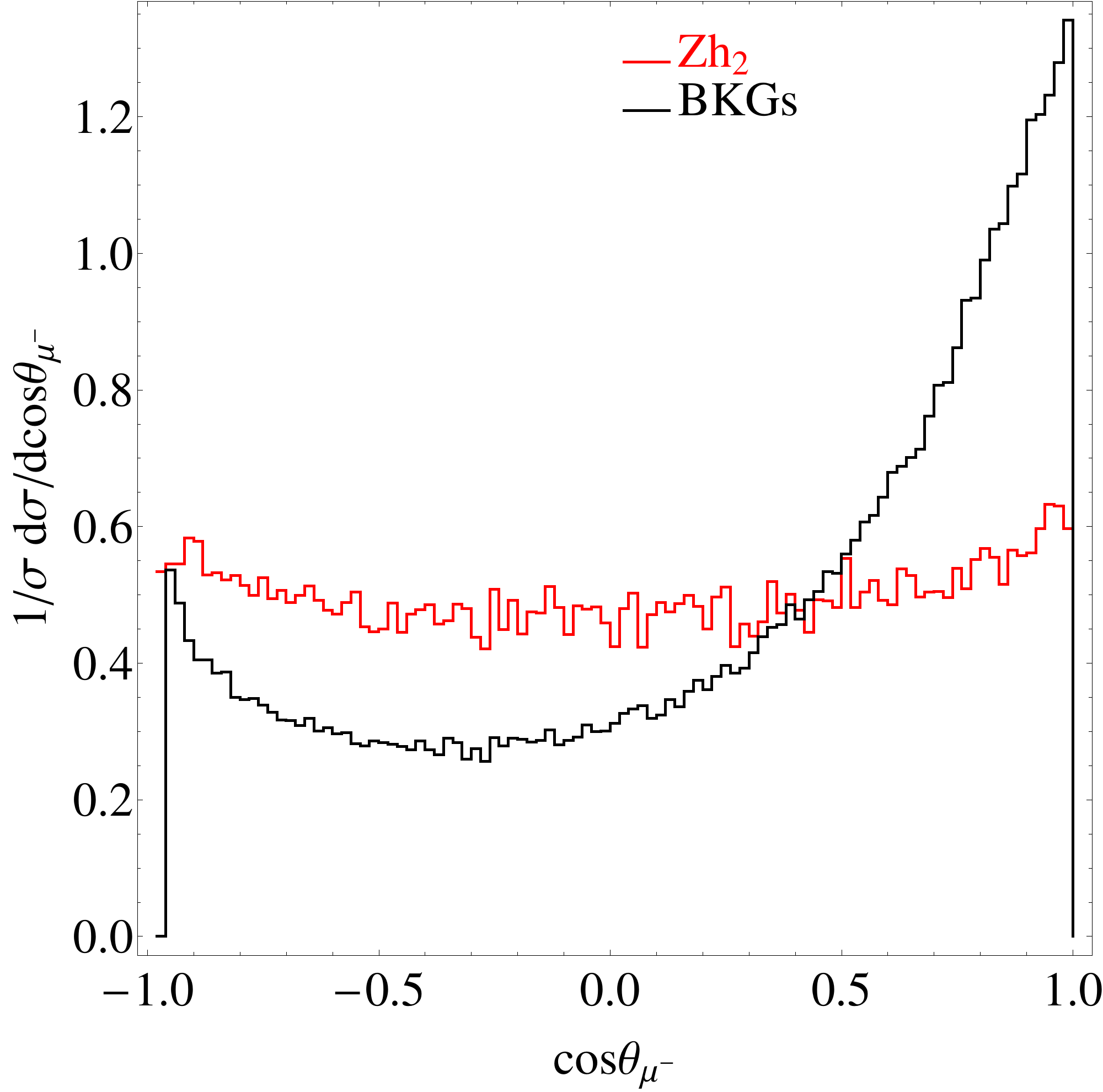}\includegraphics[scale=.19]{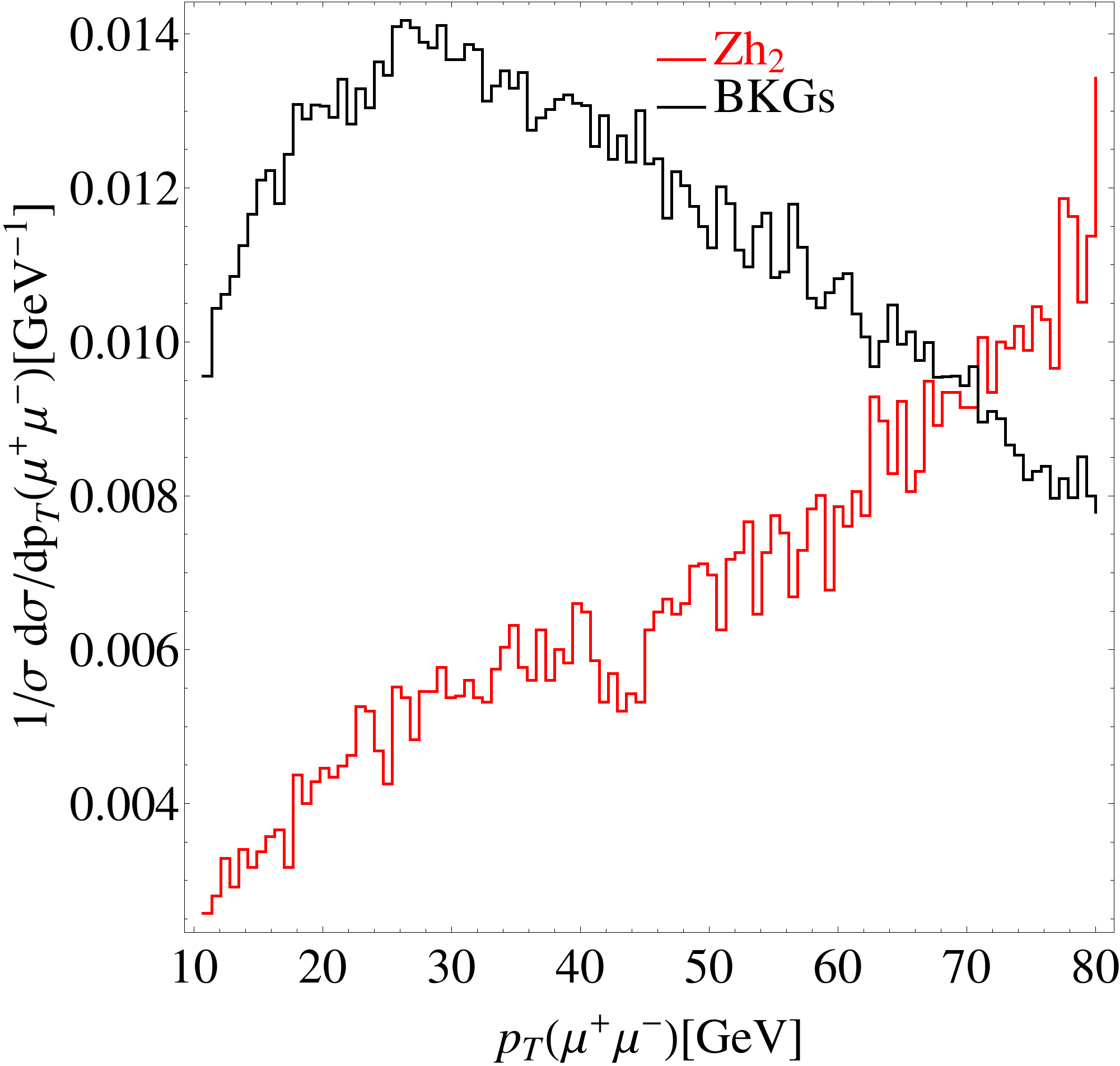}
\includegraphics[scale=.187]{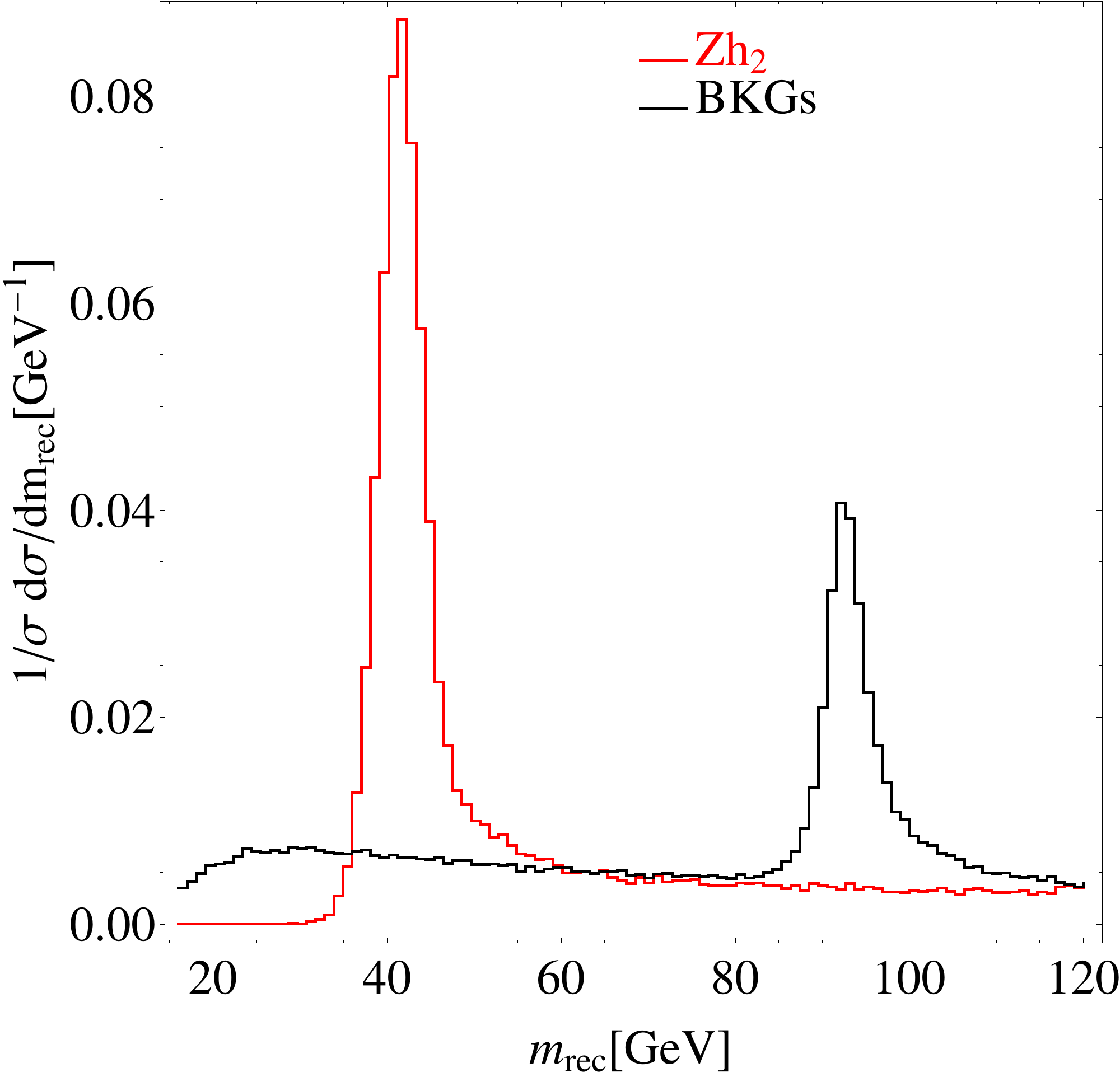}\includegraphics[scale=.187]{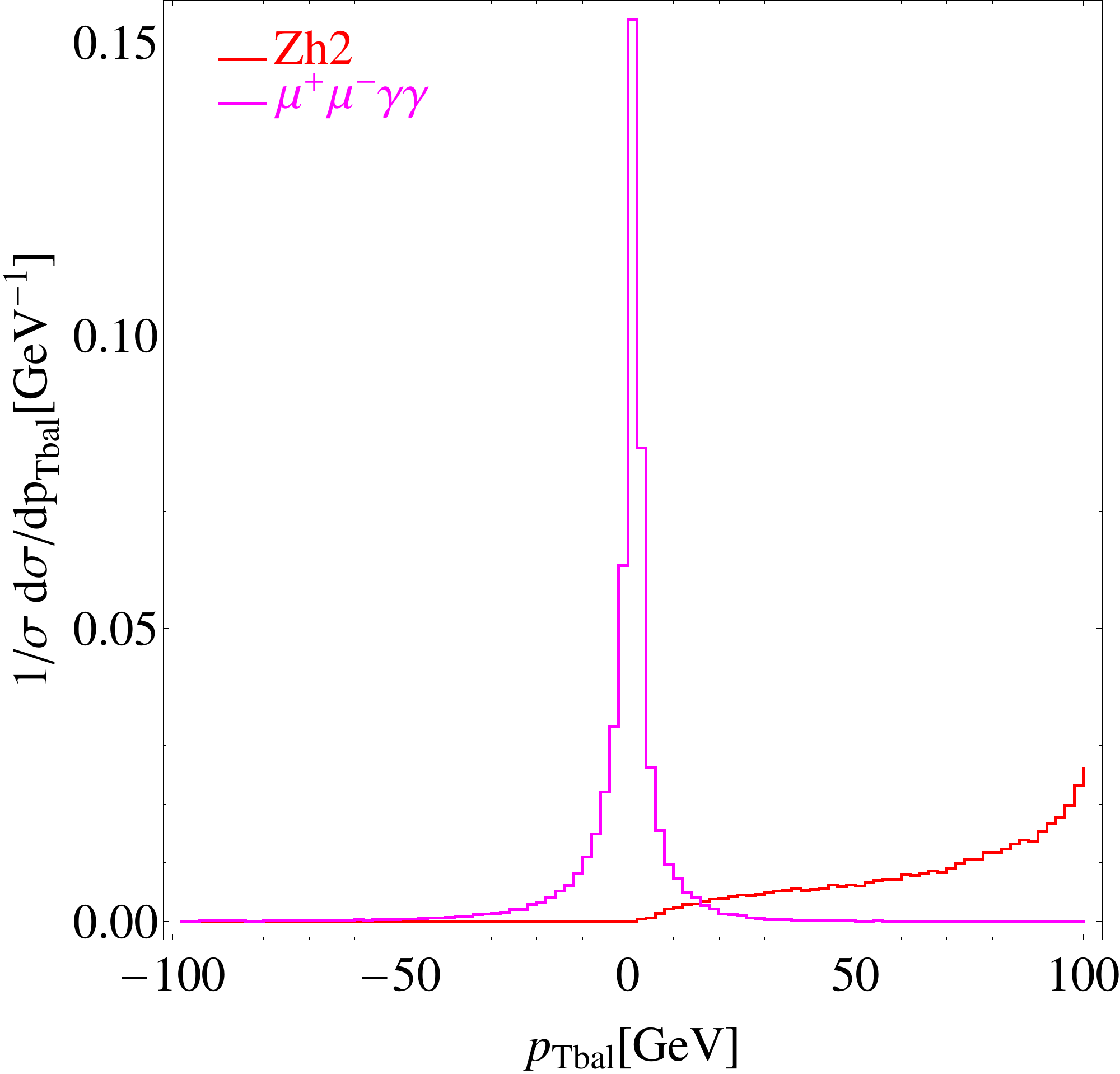}
\end{figure}
We show some kinematical distributions in \autoref{Zh2}. Based on the kinematical differences shown in the first
three figures in \autoref{Zh2}, we impose the selection cuts as
\begin{eqnarray}
&|\cos\theta_{\mu^{\pm}}|<0.8,\quad p_T(\mu^+\mu^-)>35\textrm{GeV},\quad |m_{\mu^+\mu^-}-m_Z|<10\textrm{GeV},\nonumber\\ &\textrm{and}\quad 30\textrm{GeV}<m_{\textrm{rec}}<60\textrm{GeV}.
\end{eqnarray}
The cuts on $\cos\theta_{\mu^-}$ and $p_T(\mu^+\mu^-)$ are helpful to reduce large $\mu^+\mu^-\nu\bar{\nu}$ and $\mu^+\mu^-\gamma\gamma$
backgrounds. The $m_{\mu^+\mu^-}$ cut
is imposed to extract the signal events around $Z$ peak in $m_{\mu^+\mu^-}$ distribution, and the recoil mass cut is imposed to extract
the signal events around $h_2$ peak in $m_{\textrm{rec}}$ distribution. After all the selection cuts, the cross sections of the
signal and backgrounds are
\begin{equation}
\sigma_{\textrm{sig}}=c^2_2\times7.438\textrm{fb},\quad\quad \sigma_{\textrm{bkg}}=5.916\textrm{fb},
\end{equation}
in which $e^+e^-\to \mu^+\mu^-\gamma\gamma$ is the dominant background process with the cross section $\sigma_{\mu^+\mu^-\gamma\gamma}=4.659\textrm{fb}$. Moreover, we can take advantage of the
``$p_T$ balance" cut \cite{BKG3,ptb} to suppress the $\mu^+\mu^-\gamma\gamma$ background further. The observable $p_{T,\textrm{bal}}$
is defined as
\begin{equation}
p_{T,\textrm{bal}}\equiv p_T(\mu^+\mu^-)-p_T(\gamma)
\end{equation}
where $p_T(\gamma)$ is the transverse momentum of the most energetic photon tagged \footnote{With this method, we must tag at least one
photon which breaks the inclusiveness of the measurement. But for most cases, we can assume $\textrm{Br}(h\rightarrow\gamma\gamma)\ll1$
so that tagging a photon would make only a little difference on the measurement.}. Based on the last figure in \autoref{Zh2}, if we choose
the cut $p_{T,\textrm{bal}}>20\textrm{GeV}$ as \cite{ptb}, we have
\begin{equation}
\sigma'_{\mu^+\mu^-\gamma\gamma}=0.211\textrm{fb}\quad\textrm{thus}\quad\sigma'_{\textrm{bkg}}=1.468\textrm{fb}
\end{equation}
with cross sections of other processes unchanged. Using these results, we summarize the $3\sigma$, $5\sigma$ discovery potential
and expected $95\%$ C.L. upper limit (corresponding to $1.64\sigma$) on $|c_2|$ with $5\textrm{ab}^{-1}$ luminosity
at CEPC before and after ``$p_T$ balance" cut separately in \autoref{c2}.
\begin{table}[h]
\caption{Expected $95\%$ C.L. upper limit, $3\sigma$, and $5\sigma$ discovery potential for $|c_2|$
with $5\textrm{ab}^{-1}$ luminosity at CEPC.}\label{c2}
\begin{tabular}{|c|c|c|c|}
\hline
&$95\%$ C.L. limit&$3\sigma$ discovery&$5\sigma$ discovery\\
\hline
before ``$p_T$ balance" cut&$<0.087$&$>0.118$&$>0.152$\\
\hline
after ``$p_T$ balance" cut&$<0.061$&$>0.083$&$>0.107$\\
\hline
\end{tabular}
\end{table}

For $h_1h_2$ process, we use the $h_1\rightarrow b\bar{b}$ decay channel. The backgrounds include $e^+e^-\rightarrow b\bar{b}X$
and $e^+e^-\rightarrow Zh_1(b\bar{b})$ where $X=e^+e^-$, $\mu^+\mu^-$, $\tau^+\tau^-$, $q\bar{q}$, $b\bar{b}$, $\nu\bar{\nu}$,
$\gamma\gamma$, $g\gamma$, and $gg$ \footnote{We also considered other background processes like $e^+e^-\rightarrow b\bar{b}h_{1,2}$
and $e^+e^-\rightarrow Z(b\bar{b})h_2$. However, numerically they are all negligible except for a very strong $h_{1,2}b\bar{b}$ coupling,
thus we don't list them here. Again the SM backgrounds $b\bar{b}g$ and $b\bar{b}\gamma$ can be completely removed at the parton level.}.
We impose the basic cuts as
\begin{eqnarray}
&m_{b\bar{b}}>15\textrm{GeV},\quad m_{\textrm{rec}}>15\textrm{GeV}, \nonumber\\
&|\cos\theta_{e^{\pm},\gamma}|<0.995,\quad E_{\gamma}>0.1\textrm{GeV},\quad E_g>1\textrm{GeV},\quad \Delta R_{ij}>0.4
\end{eqnarray}
where $m_{\textrm{rec}}$ is defined in (\ref{rec}) with $f=b$ and $\Delta R_{ij}$ run over all partons in the final
state \footnote{The cuts in the second line are useful to avoid the infrared and collinear divergences in background processes
as discussed above.}. The jet energy is smeared by a Gaussian distribution with the standard deviation of \cite{CEPC}
\begin{equation}
\frac{\sigma_E}{E}=\frac{0.3}{\sqrt{E(\textrm{GeV})}}
\end{equation}
for the jet energy less than $100\textrm{GeV}$. The $b$-tagging efficiency and $c$-faking rate are \cite{CEPC}
\begin{equation}
\epsilon_b=0.9,\quad\quad P_{c\rightarrow b}=0.1
\end{equation}
separately. In an event, at least two $b$ jets should be tagged. The candidates of $b$ jets from $h_1$ decays are selected
with the minimal $|m_{b\bar{b}}-m_1|$ and then sorted by the transverse momenta. The leading and sub-leading $p_T$ of the
selected $b$ jet pairs are denoted as $p_T(b)$ and $p^{\textrm{sub}}_T(b)$.

\begin{figure}[h]
\caption{Normalized kinematical distributions of the signal and backgrounds in the $e^+e^-\rightarrow h_1h_2$ channel after the
basic cuts are applied and $\geq2b$ jets are tagged. In the first five figures, only $b$ jets are reconstructed; while in the last
figure, at least one photon should be tagged which breaks the inclusiveness a little bit.}\label{h1h2}
\includegraphics[scale=.255]{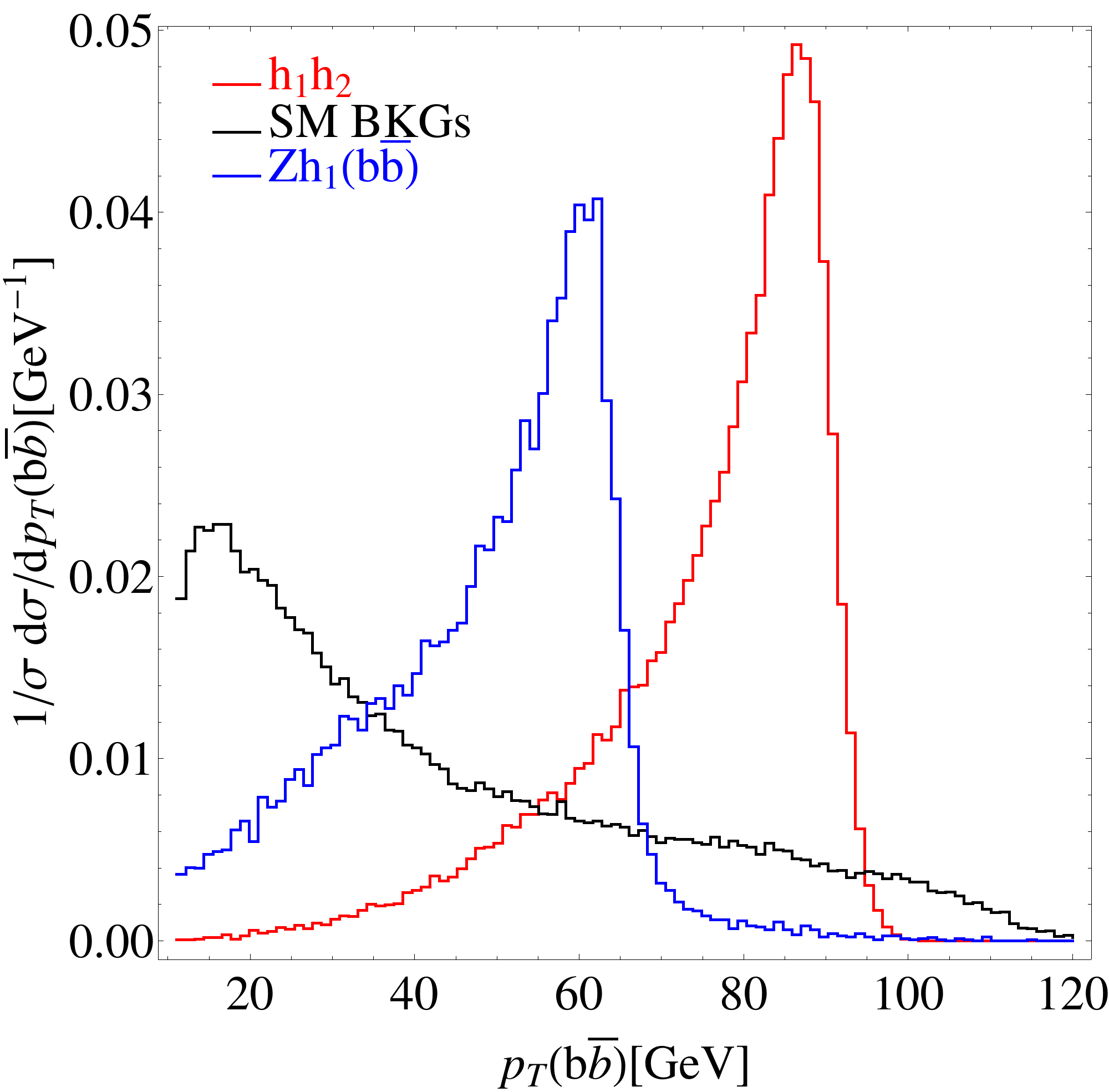}\includegraphics[scale=.26]{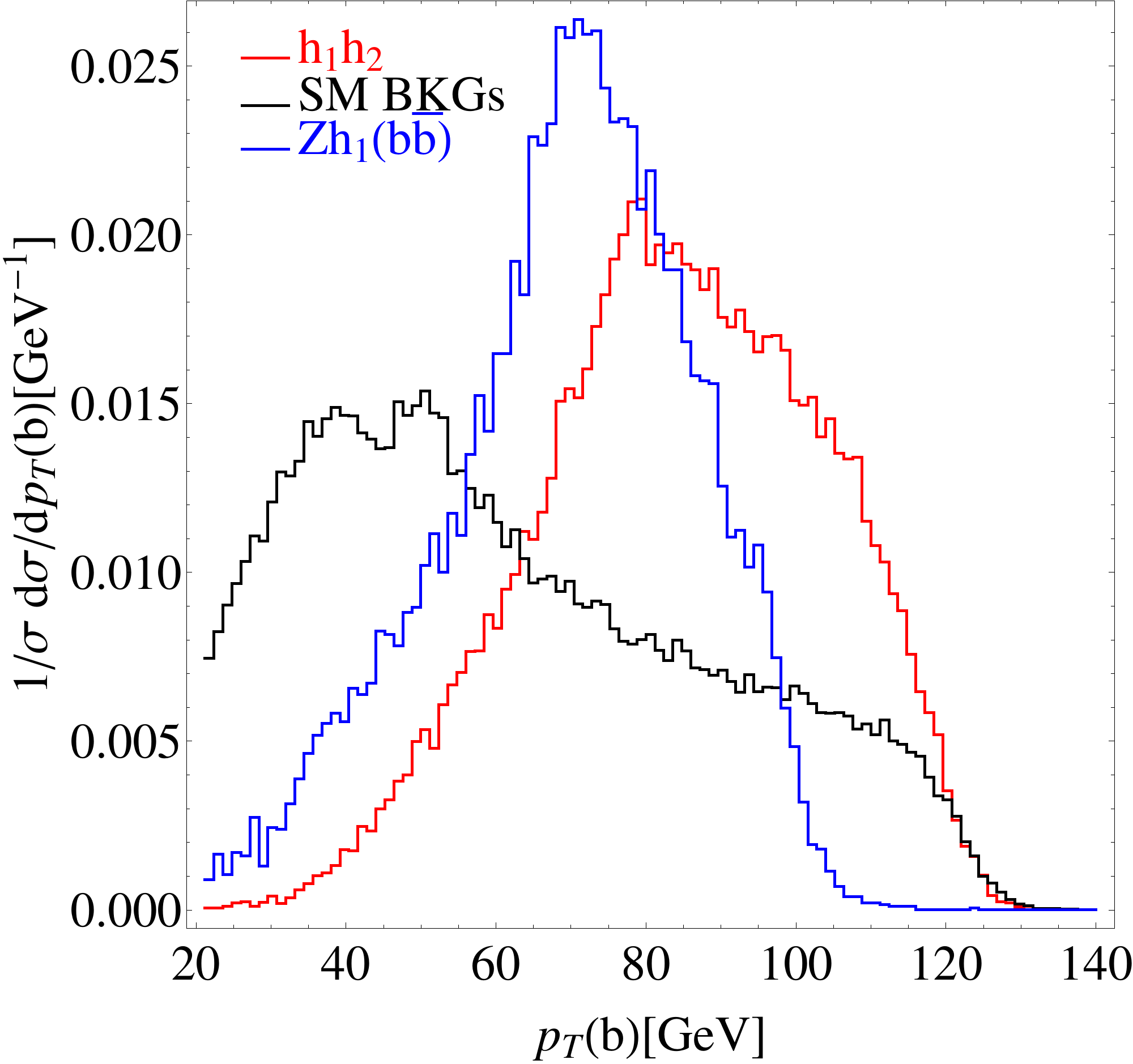}\includegraphics[scale=.25]{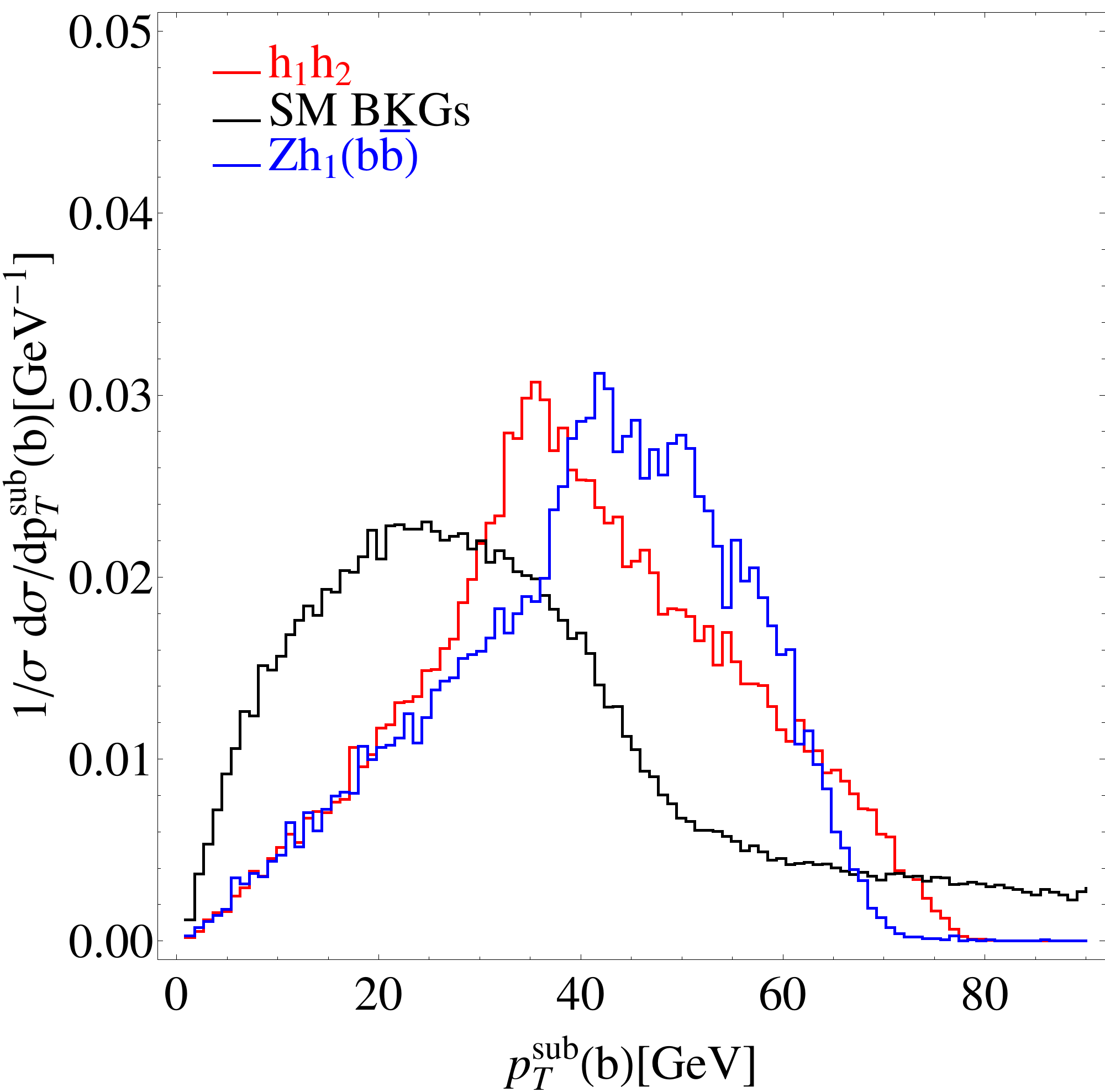}
\includegraphics[scale=.26]{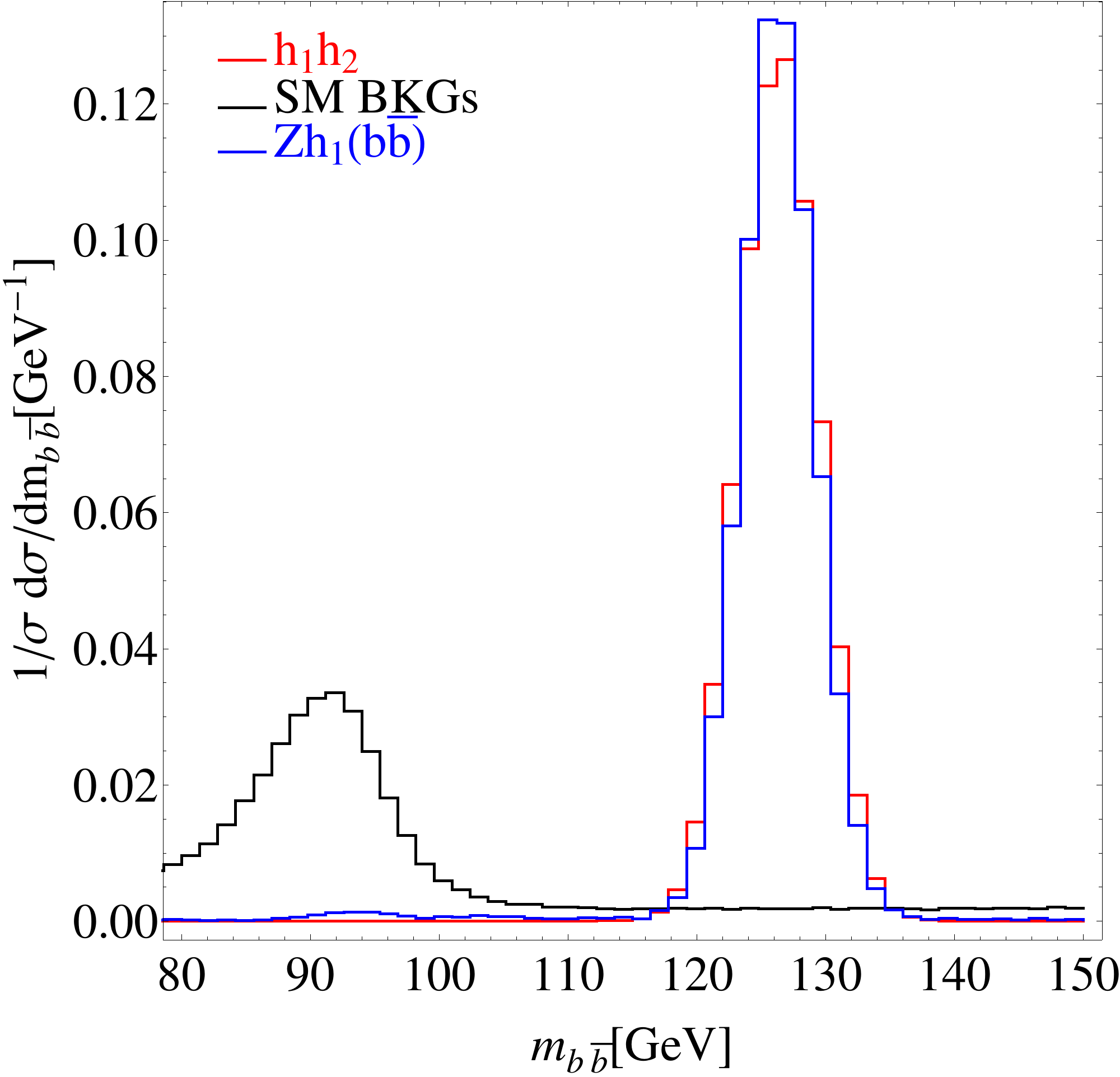}\includegraphics[scale=.26]{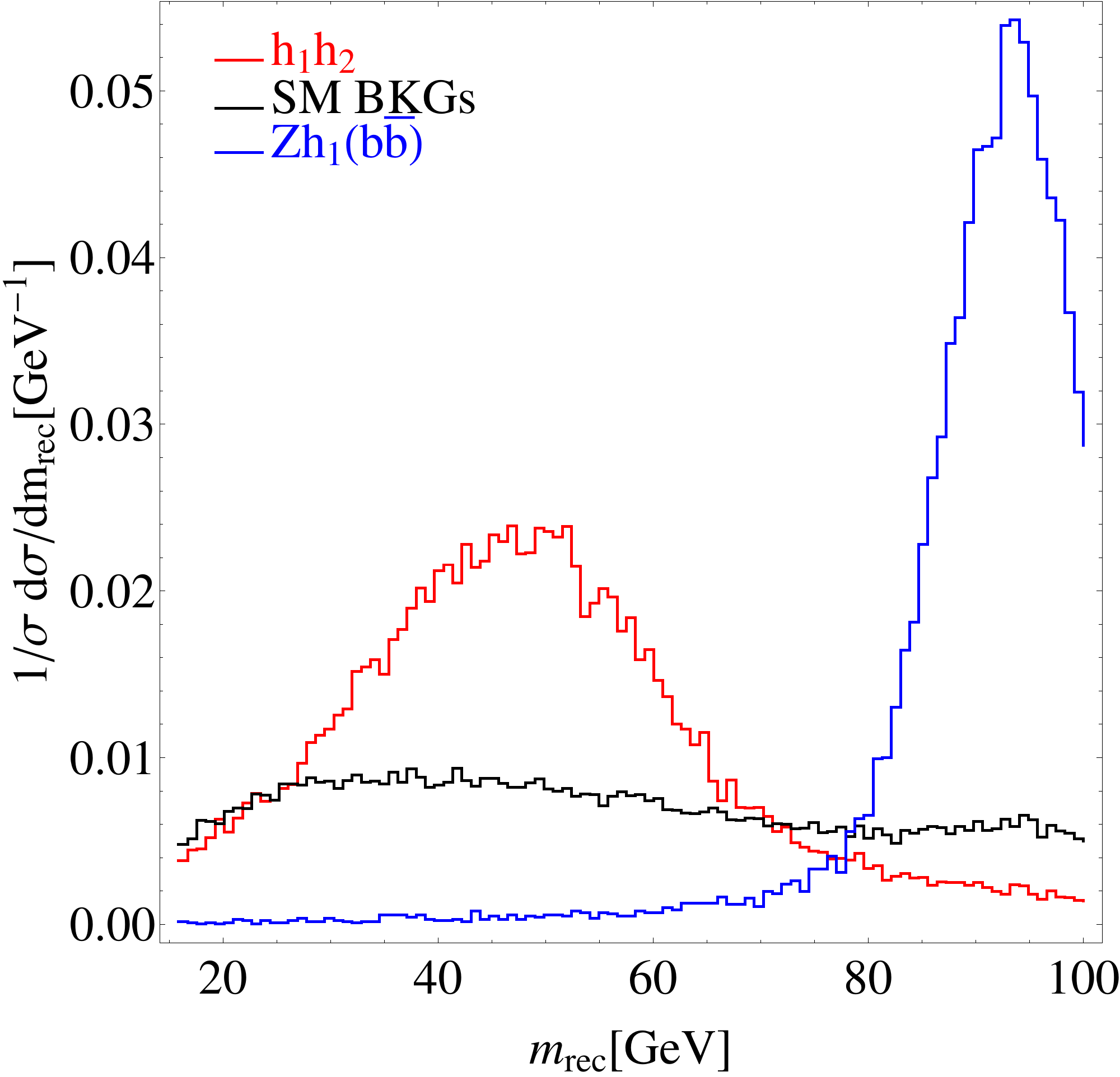}\includegraphics[scale=.26]{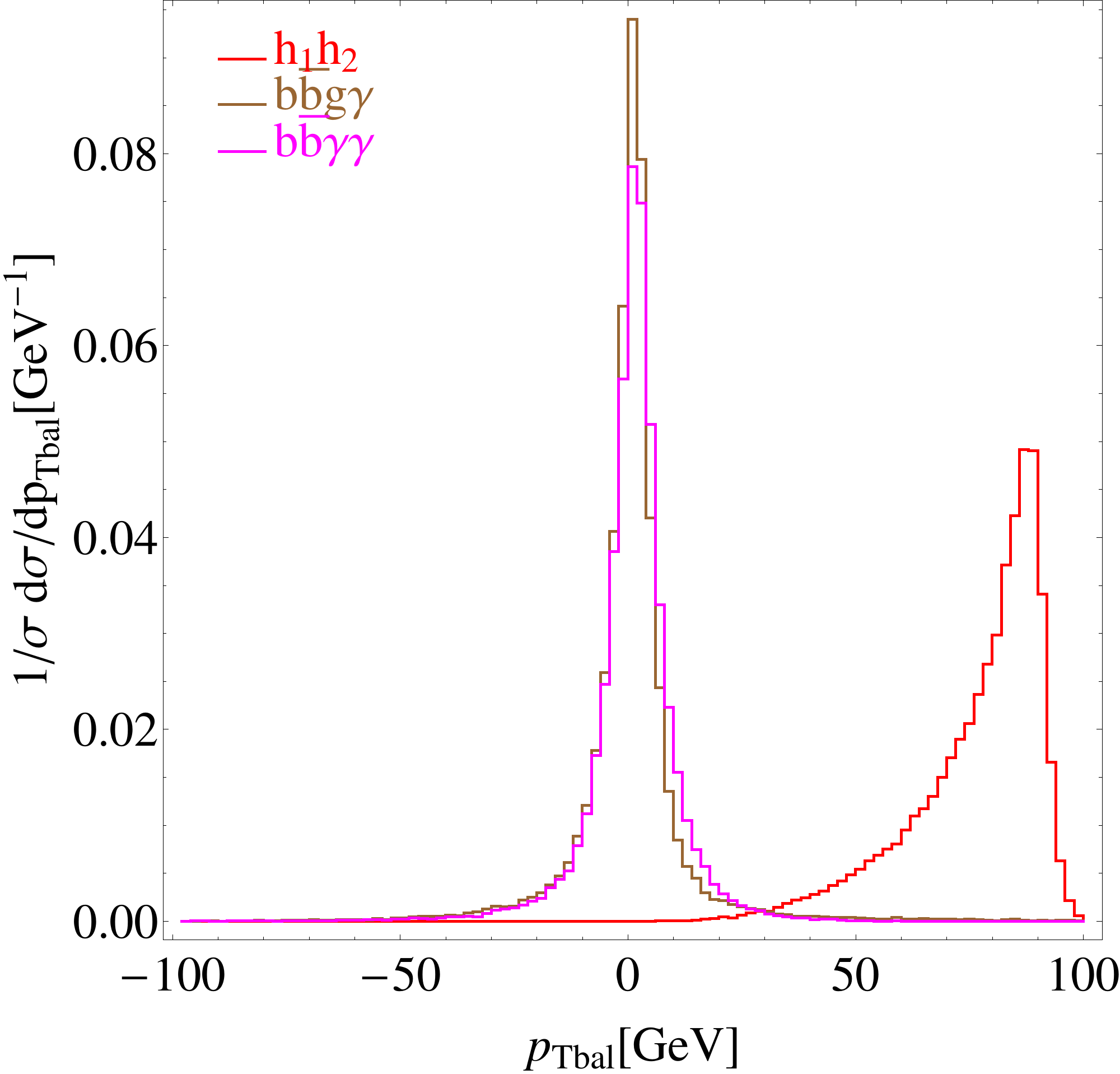}
\end{figure}
Based on the kinematic differences between the signal and backgrounds as shown in the first five figures in \autoref{h1h2},
we impose the selection cuts as
\begin{eqnarray}
&70\textrm{GeV}<p_T(b\bar{b})<100\textrm{GeV},\quad70\textrm{GeV}<p_T(b)<110\textrm{GeV},\quad30\textrm{GeV}<p^{\textrm{sub}}_T(b)<70\textrm{GeV},\nonumber\\
&\quad |m_{b\bar{b}}-m_1|<25\textrm{GeV},\quad\textrm{and}\quad20\textrm{GeV}<m_{\textrm{rec}}<70\textrm{GeV}.
\end{eqnarray}
The cuts in the first line use the differences in $b$ jets $p_T$ distributions to distinguish events from signal and backgrounds.
The $m_{b\bar{b}}$ cut is imposed to extract the signal events around $Z$ peak in $m_{b\bar{b}}$ distribution, and the recoil mass
cut is imposed to extract the signal events around $h_2$ peak in $m_{\textrm{rec}}$ distribution. After these selection cuts,
the cross sections of signal and backgrounds are
\begin{equation}
\sigma_{\textrm{sig}}=\frac{c^2_{12}\textrm{Br}(h_1\rightarrow b\bar{b})}{\textrm{Br}_{\textrm{SM}}(h_1\rightarrow b\bar{b})}\times12.5\textrm{fb},\quad
\sigma_{\textrm{bkg}}=\left(20.54+0.577\left(\frac{\textrm{Br}(h_1\rightarrow b\bar{b})}{\textrm{Br}_{\textrm{SM}}(h_1\rightarrow b\bar{b})}\right)\right)\textrm{fb}
\end{equation}
where $\textrm{Br}_{\textrm{SM}}(h_1\rightarrow b\bar{b})=0.5824$ \cite{Br} for $m_1=125\textrm{GeV}$. The dominant background is
$b\bar{b}gg$ production with its cross section $\sigma_{b\bar{b}gg}=13.2\textrm{fb}$. The backgrounds with photon have the cross section
$\sigma_{b\bar{b}g\gamma+b\bar{b}\gamma\gamma}=4.981\textrm{fb}$ which can be suppressed to $\sigma'_{b\bar{b}g\gamma+b\bar{b}\gamma\gamma}
=1.107\textrm{fb}$ by using the ``$p_T$ balance" cut based on the last figure in \autoref{h1h2}. The ``$p_T$ balance"
cut does not affect the signal and other background processes thus the total background can be reduced to
\begin{equation}
\sigma'_{\textrm{bkg}}=\left(16.66+0.577\left(\frac{\textrm{Br}(h_1\rightarrow b\bar{b})}{\textrm{Br}_{\textrm{SM}}(h_1\rightarrow b\bar{b})}\right)\right)\textrm{fb}.
\end{equation}
As a benchmark point, take $\textrm{Br}(h_1\rightarrow b\bar{b})=\textrm{Br}_{\textrm{SM}}(h_1\rightarrow b\bar{b})$. We use the results
above to summarize the $3\sigma$, $5\sigma$ discovery potential and expected $95\%$ C.L. upper limit on $|c_{12}|$ with $5\textrm{ab}^{-1}$
luminosity at CEPC before and after ``$p_T$ balance" cut separately in \autoref{c12}.
\begin{table}[h]
\caption{Expected $95\%$ C.L. upper limit, $3\sigma$, and $5\sigma$ discovery potential for $|c_{12}|$
with $5\textrm{ab}^{-1}$ luminosity at CEPC.}\label{c12}
\begin{tabular}{|c|c|c|c|}
\hline
&$95\%$ C.L. limit&$3\sigma$ discovery&$5\sigma$ discovery\\
\hline
before ``$p_T$ balance" cut&$<0.092$&$>0.125$&$>0.161$\\
\hline
after ``$p_T$ balance" cut&$<0.088$&$>0.119$&$>0.153$\\
\hline
\end{tabular}
\end{table}

For $m_2<125\textrm{GeV}$, the three processes $e^+e^-\rightarrow Zh_1,Zh_2,h_1h_2$ are possible at CEPC. However, the
method discussed in this paper is not always effective for the whole mass region. If $m_2\lesssim34\textrm{GeV}$ when rare
decay $h_1\rightarrow Zh_2$ process opens, it will set a stricter constraint $|c_{12}|\lesssim0.07$ which make this
method invalid \cite{our2}. For a larger $m_2$, both cross sections $\sigma_{Zh_2,h_1h_2}$ decrease when $m_2$ grows. But when $m_2$
is not close to $Z$ peak, for example, $m_2\lesssim70\textrm{GeV}$, the cross sections of signal and backgrounds change
slowly thus the method will still be effective. For example, when $m_2=70\textrm{GeV}$, our simulations show that the
$5\sigma$ discovery bound can reach $|c_2|>0.13(0.11)$ and $|c_{12}|>0.21(0.20)$ respectively before (after) ``$p_T$ balance" cut.
For $m_2\sim(70-110)\textrm{GeV}$ which is around the $Z$ peak, large $Z$ backgrounds will be difficult to reduce for both
$e^+e^-\rightarrow Zh_2,h_1h_2$ which means the analysis we used above is not enough and we may need more careful analysis.
For larger $m_2$, the $h_1h_2$ production cross section will decrease quickly when $m_2$ grows. Thus at CEPC, this method is
effective for $m_2\sim(35-70)\textrm{GeV}$.

\section{Implication for Weakly-Coupled Lee Model}
\label{leemodel}
In this paper, we the choose weakly-coupled Lee model
\cite{our2,thesis} which naturally contains a light scalar in small CP-violation limit as a benchmark model
to study the implications of our simulation results.

Lee model was proposed by Lee in 1973 \cite{Lee} as a 2HDM which is CP-conserved at Lagrangian level but the CP-violation
comes from the vacuum. The scalar potential can be written as
\begin{eqnarray}
V(\phi_1,\phi_2)&=&\mu_{1}^2R_{11}+\mu_{2}^2R_{22}+\lambda_1R_{11}^2+\lambda_2R_{11}R_{12}\nonumber\\
&&+\lambda_3R_{11}R_{22}+\lambda_4R_{12}^2+\lambda_5R_{12}R_{22}+\lambda_6R_{22}^2+\lambda_7I_{12}^2
\end{eqnarray}
where $R(I)_{ij}$ is the real (imaginary) part of $\phi_i^{\dag}\phi_j$. Both $\phi_i$ are scalar doublets which can be
written as $\phi_1=(\phi_1^+,(v_1+R_1+\textrm{i}I_1)/\sqrt{2})^T$ and $\phi_2=(\phi_2^+,(v_2\exp(\textrm{i}\xi)+R_2+\textrm{i}I_2)
/\sqrt{2})^T$. Here $I_{1,2}$ and $R_{1,2}$ are scalar degrees of freedom and $v=\sqrt{v_1^2+v^2_2}=246\textrm{GeV}$. According to
the vacuum stability condition, if
\begin{equation}
|\lambda_2v_1^2+\lambda_5v^2_2|<2|\lambda_4-\lambda_7|v_1v_2,
\end{equation}
a nontrivial phase difference $\xi$ between the vacuum expected values (VEV) of the two Higgs doublets would arise thus CP symmetry
is spontaneously broken. As a consequence all the three neutral Higgs bosons must be CP-mixing states.

Defining $t_{\beta}\equiv v_2/v_1$, for weakly-coupled scalar sector ($\lambda_i\lesssim\mathcal{O}(1)$), in the limit of small
$t_{\beta}s_{\xi}$, a new light scalar is predicted with the mass $m_2\sim\mathcal{O}(vt_{\beta}s_{\xi})$ \cite{our1,our2,thesis}. We treat it as the
$40\textrm{GeV}$ new scalar. Its couplings to massive vector bosons are also suppressed by $c_2\sim\mathcal{O}(t_{\beta}s_{\xi})
\sim\mathcal{O}(0.1)$. If the heavy Higgs boson has its mass $m_3\sim\mathcal{O}(v)$, there is also additional constraint on
$c_{12}$ from LHC results \cite{recent}. If $200\textrm{GeV}<m_3<300\textrm{GeV}$, $c_{12}\equiv c_3\lesssim(0.3-0.4)$ \cite{our2,thesis,talk}
which is stricter than the LEP result. In this scenario, the $125\textrm{GeV}$ Higgs boson $h_1$ has SM-like couplings. The $h_1\rightarrow2h_2$
decay channel measurements impose a strict constraint on $h_1h_2h_2$ coupling to $\mathcal{O}(10^{-2})$ \cite{our2,LHCrare}, but this measurement
does not give tighter constraints on the $c_1$, $c_2$ and $c_{12}$ couplings. The electro-weak precision measurements \cite{EWP} require that the
charged Higgs boson mass must be close the the heavy Higgs mass $m_3$ \cite{our2}. For $m_3\sim v$, there is no further constraints from $t\rightarrow
H^+b$ rare decay \cite{our2}. The study in \cite{our2} and its update results in \cite{talk} showed this scenario is still viable facing all experimental constraints.

\begin{figure}[h]
\caption{Expected limit or significance for different $(c_2,c_{12})$ points. The left figure is for the results before ``$p_T$ balance"
cut while the right figure is for the results after ``$p_T$ balance" cut thus it is a ``quasi-inclusive" result. The four curves are for
$K=0.01,0.02,0.03,0.04$ from left to right. We denote the boundary of $1.64\sigma$, $3\sigma$, and $5\sigma$ significance with green, blue,
and cyan lines respectively.}\label{LM}
\includegraphics[scale=.6]{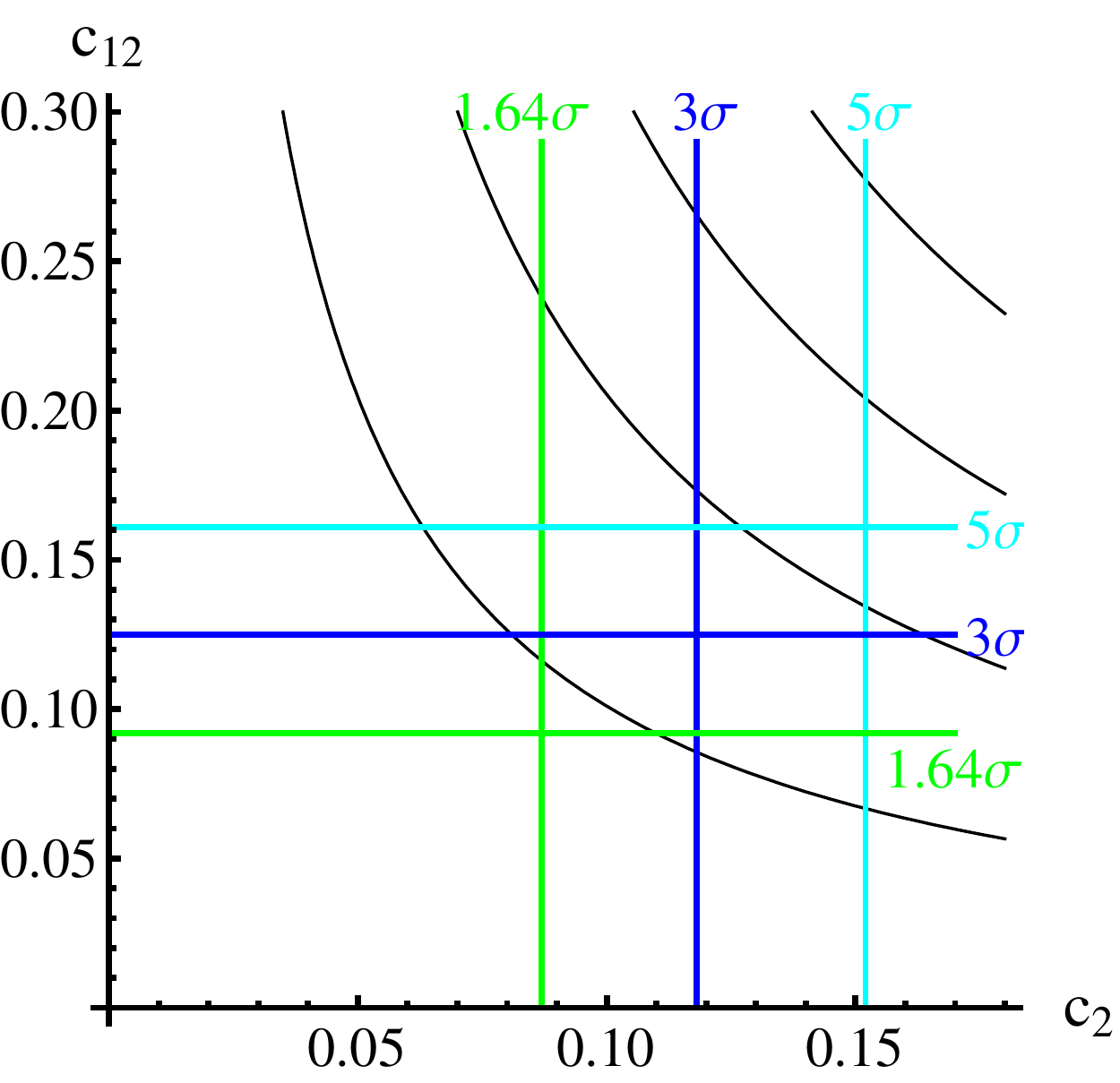}\includegraphics[scale=.6]{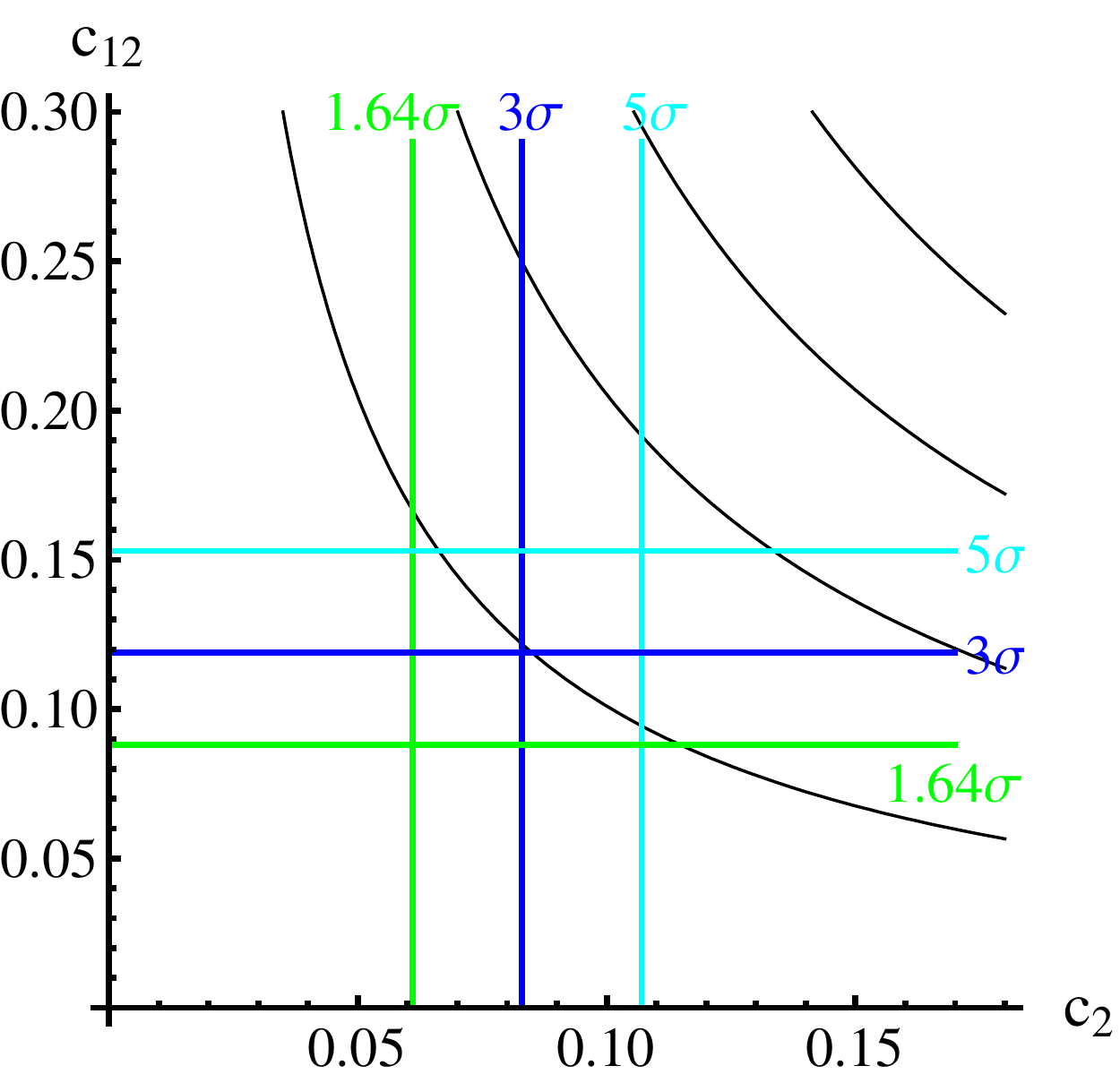}
\end{figure}

The results we obtained above showed we can set stricter constraint or discovery potential on this scenario. For $h_1h_2$ production channel,
we use $\textrm{Br}(h_1\rightarrow b\bar{b})=\textrm{Br}_{\textrm{SM}}(h_1\rightarrow b\bar{b})$ as a benchmark point.
Assuming all $c_{1,2,12}>0$, we have
\begin{equation}
K=c_2c_{12}\sqrt{1-c^2_2-c^2_{12}}.
\end{equation}
In \autoref{LM} we show the expected limit or significance for different $(c_2,c_{12})$ points before (see the left figure) or after
(see the right figure) ``$p_T$ balance" cut discussed above. The four curves are $K=0.01,0.02,0.03,0.04$ separately from left to right.

If there is no hint for either processes before ``$p_T$ balance" cut, it is expected to set an upper limit $K<7.9\times10^{-3}$;
while the upper limit is expected to be $K<5.3\times10^{-3}$ after ``$p_T$ balance" cut. If both processes are discovered at over $3(5)\sigma$
level before ``$p_T$ balance" cut, we have $K>1.5(2.4)\times10^{-2}$; while the number should be $1.0(1.6)\times10^{-2}$ after ``$p_T$
balance" cut. In this case, we can confirm CP-violation in the scalar sector and measure $K$ to the accuracy $\delta K/K\lesssim24(16)\%$.
For the case with the largest $K$, both couplings are set to the recent allowed upper limit, $c_2=0.18$ and $c_{12}=0.3$, we will have $K=
5.4\times10^{-2}$ and $\delta K/K=7.9(4.7)\%$ before (after) ``$p_T$ balance" cut.

In the discussions above, we just use the inclusive measurements to determine the couplings and hence $K$ in a model-independent way.
For the discovery potential of a specific model, it would be better to use exclusive decay channels such as $h_2\rightarrow b\bar{b}$
which is expected to be dominant in most models. The sensitivity would also increase if we combine the results from more decay channels
of $Z$ and $h_1$.

\section{Conclusions and Discussions}
\label{conc}
Once two scalars are discovered, we can test the CP-violation in the scalar sector through searching for nonzero
tree-level $h_1ZZ,h_2ZZ,h_1h_2Z$ vertices according to the CP-properties analysis. Based on this idea, we proposed
a model-independent method to confirm CP-violation in the scalar sector at future $e^+e^-$ colliders through measuring
the inclusive $e^+e^-\rightarrow Zh_1,Zh_2,h_1h_2$ cross sections with recoil mass technique. We can
use a quantity $K=c_1c_2c_{12}$ which is defined in (\ref{kdef}) to measure CP-violation in the scalar sector.

We have performed simulation studies for $m_2=40\textrm{GeV}$ at CEPC assuming the $125\textrm{GeV}$ Higgs boson $h_1$ is SM-like and
the results are shown in \autoref{c2} and \autoref{c12}. We have adopted the recoil mass technique to ensure the measurements
are inclusive \footnote{After ``$p_T$ balance" cut, it is quasi-model-independent as
discussed above.}. The $5\sigma$ discovery limit for both $c_2$ and $c_{12}$
are below the recent $95\%$ C.L. upper limits. For $Zh_2$ associated production, the ``$p_T$ balance" cut
is efficient to drop the photon background but it also lose the inclusiveness a little.
We choose the weakly-coupled
Lee model which contains CP-violation and allows an extra light scalar as a benchmark model \footnote{For most perturbative models, this
method is useful to extract tree level information instead of loop level since loop-induced processes have small enough cross sections.}.
In the weakly coupled Lee model,
both processes, $Zh_2$ and $h_1h_2$, are possible to be discovered at $5\sigma$ level before or after ``$p_T$ balance" cut.
If both processes are discovered at $3(5)\sigma$ level, $K$ must reach $\mathcal{O}(10^{-2})$ and the sensitivity of $\delta K/K$
measurement can reach $24(16)\%$. This method is also applicable for other $e^+e^-$ colliders if all the three processes can be
discovered. For example, if the extra scalar is heavier, we can use this method at a $e^+e^-$ collider with larger $\sqrt{s}$, such as ILC.

We should note that $K\neq0$ is a sufficient but not necessary condition for the existence of CP-violation in the scalar
sector. Precisely speaking, we can use this method to confirm the existence of CP-violation in the scalar sector according to
the nonzero $K$, but can't constrain or exclude the CP-violation in the scalar sector if $K$ is unmeasurable small.
For example, in a minimal extension of SM mentioned above \cite{singlet}, there is only an additional complex
singlet in the extension of the scalar sector. For some parameter choices, the three scalars would become CP-mixing states,
but there are still no tree-level $h_ih_jZ$ vertices thus the measurement on $e^+e^-\rightarrow h_ih_j$ cannot give a
positive result.

\section*{Acknowledgement}
We thank Yacine Haddad, Wolfgang Kilian, Gang Li (IHEP), Zhen Liu, Xin Mo, Juergen Reuter, Man-Qi Ruan, Yi-Lei Tang, Xia Wan,
and Hao Zhang for helpful discussions. This work was supported in part by the Natural Science Foundation of China (Grants No. 11635001,
No. 11135003, and No. 11375014).

\numberwithin{equation}{section}

\clearpage\end{CJK*}


\begin{thebibliography}{99}

\bibitem{CPVdisc}J. H. Christenson, J. W. Cronin, V. L. Fitch, and R. Turlay, Phys. Rev. Lett. 13, 138 (1964).
\bibitem{PDG}K. A. Olive et. al. (Particle Data Group), Chin. Phys. C 38, 090001 (2014); Chin. Phys. C 40, 100001 (2016).
\bibitem{KM}M. Kobayashi and T. Maskawa, Prog. Theor. Phys. 49, 652 (1973).
\bibitem{CKM}N. Cabibbo, Phys. Rev. Lett. 10, 531 (1963).
\bibitem{Plank}The Planck Collaboration, Astron. Astrophys. 571, A16 (2014).
\bibitem{2HDM}G. C. Branco, P. M. Ferreira, L. Lavoura, M. N. Rebelo, M. Sher, and J. P. Silva, Phys. Rep. 516, 1 (2012).
\bibitem{singlet}L. Bento, G. C. Branco, and P. A. Parada, Phys. Lett. B 267, 95 (1991).
\bibitem{Lee}T. D. Lee, Phys. Rev. D 8, 1226 (1973); Phys. Rep. 9, 143 (1974).
\bibitem{georgi}H. Georgi, Hadronic J. 1, 155 (1978).
\bibitem{3HDM}S. Weinberg, Phys. Rev. Lett. 37, 657, (1976).
\bibitem{our1}Y.-N. Mao and S.-H. Zhu, Phys. Rev. D 90, 115024 (2014).
\bibitem{our2}Y.-N. Mao and S.-H. Zhu, Phys. Rev. D 94, 055008 (2016); Phys. Rev. D 94, 059904 (2016, erratum added).
\bibitem{thesis}Y.-N. Mao, PhD Thesis (Peking University, 2016).
\bibitem{ATLAS}The ATLAS Collaboration, Phys. Lett. B 716, 1 (2012).
\bibitem{CMS}The CMS Collaboration, Phys. Lett. B 716, 30 (2012).
\bibitem{higgsmass}The ATLAS and CMS Collaborations, Phys. Rev. Lett. 114, 191803, (2015).
\bibitem{spinCP1}The CMS Collaboration, Phys. Rev. D 89, 092007 (2014).
\bibitem{spinCP2}The CMS Collaboration, Report No. CMS-PAS-HIG-14-014. 	
\bibitem{spinCP3}The ATLAS Collaboration, Report No. ATLAS-CONF-2015-008.
\bibitem{spinCP4}The CMS Collaboration, Reports No. CMS-HIG-14-035 and No. CERN-PH-EP/2015-331, arXiv: 1602.00209.
\bibitem{hsc}L. Bian and N. Chen, J. High Energy Phys. 09, 069 (2016).
\bibitem{EDM}M. Pospelov and A. Ritz, Ann. Phys. (Amsterdam) 318, 119 (2005).
\bibitem{eEDM}ACME Collaboration, Science 343, 269 (2014).
\bibitem{nEDM}C. A. Baker et. al., Phys. Rev. Lett. 97, 131801 (2006); J. M. Pendlebury et. al., Phys. Rev. D 92, 092003 (2015).
\bibitem{mixing}A. Hocker and Z. Ligeti, Ann. Rev. Nucl. Part. Sci. 56, 501 (2006); J. Charles, S. Descotes-Genon, Z. Ligeti,
 S. Monteil, M. Papucci, and K. Trabelsi, Phys. Rev. D 89, 033016 (2014).
\bibitem{GH}J. F. Gunion and H. E. Haber, Phys. Rev. D 72, 095002 (2005).
\bibitem{ZZZ1}B. Grzadkowski, O. M. Ogreid, and P. Osland, J. High Energy Phys. 11, 084 (2014); PoS CORFU2014, 086 (2015).
\bibitem{ZZZ2}B. Grzadkowski, O. M. Ogreid, and P. Osland, J. High Energy Phys. 05, 025 (2016).
\bibitem{CPtau1}S. Berge, W. Bernreuther, and J. Ziethe, Phys. Rev. Lett. 100, 171605 (2008).
\bibitem{CPtau2}S. Berge, W. Bernreuther, and S. Kirchner, Phys. Rev. D 92, 096012 (2015).
\bibitem{CPtau3}S. Berge, W. Bernreuther, and H. Spiesberger, Phys. Lett. B 727, 488 (2013).
\bibitem{CPt}P. S. Bhupal Dev, A. Djouadi, R. M. Godbole, M. M. M$\ddot{\textrm{u}}$hlleitner, and S. D. Rindani, Phys. Rev. Lett. 100, 051801 (2008).
\bibitem{K}A. M$\acute{\textrm{e}}$ndez and A. Pomarol, Phys. Lett. B 272, 313 (1991).
\bibitem{hVV}The CMS Collaboration, J. High Energy Phys. 01, 096 (2014); The ATLAS Collaboration,
 Report No. ATLAS-CONF-2015-007; The ATLAS and CMS Collaborations, Report No. ATLAS-CONF-2015-044.
\bibitem{loop}G. Li, Y.-N. Mao, Y.-L. Tang, C. Zhang, Y. Zhou, and S.-H. Zhu, Phys. Rev. Lett. 116, 151803 (2016).
\bibitem{softCP}C.-Y. Chen, S. Dawson, and Y. Zhang, J. High Energy Phys. 06, 056 (2015).
\bibitem{csx1}The ALEPH, DELPHI, L3, and OPAL Collaborations (LEP Higgs Working Group), Eur. Phys. J. C 47, 547 (2006).
\bibitem{csx2}S. Heinemeyer and C. Schappacher, Eur. Phys J. C 76, 220 (2016).
\bibitem{rec1}The NLC ZDR Design Group and NLC Physics Working Group Collaborations, Reports No. SLAC-R-0485, No. SLAC-R-485,
 No. SLAC-0485, No. SLAC-485, No. BNL-52502, No. FERMILAB-PUB-96-112, No. LBL-PUB-5425, No. LBNL-PUB-5425, No. UCRL-ID-124160,
 No. BNL-52-502, No. SLAC-REPORT-485, No. --UCRL-ID-124160, and No. UC-414, arXiv: hep-ex/9605011.
\bibitem{rec2}J. F. Gunion, T. Han, and R. Sobey, Phys. Lett. B 429, 79 (1998).
\bibitem{CEPC}The CEPC-SPPC Study Group, Reports No. IHEP-CEPC-DR-2015-01, No. IHEP-TH-2015-01, and No. HEP-EP-2015-01,
 \url{http://cepc.ihep.ac.cn/preCDR/volume.html}.
\bibitem{TLEP}The TLEP Design Study Working Group, J. High Energy Phys. 01, 164 (2014).
\bibitem{rec3}M. Thomson,  Eur. Phys. J. C 76, 72 (2016).
\bibitem{ILC}C. Adolphsen et. al., arXiv: 1306.6328; arXiv: 1306.6353.
\bibitem{2HDM1}J.-M. G$\acute{\textrm{e}}$rard and M. Herquet, Phys. Rev. Lett. 98, 251802 (2007); B. Coleppa, F. Kling, and S. Su,
 J. High Energy Phys. 01, 161 (2014).
\bibitem{2HDM2}B. Dumont, J. F. Gunion, Y. Jiang, and S. Kraml, Phys. Rev. D 90, 035021 (2014); J. Bernon, J. F. Gunion,
 Y. Jiang, and S. Kraml, Phys. Rev. D 91, 075019 (2015).
\bibitem{recent}The ATLAS collaboration, Report No. ATLAS-CONF-2016-081; the CMS collaboration, Report No. CMS-PAS-HIG-16-033.
\bibitem{LEP1}G. Abbiendi et. al. (ALEPH, DELPHI, L3, and OPAL Collaborations and the LEP Higgs Working Group),
 Phys. Lett. B 565, 61 (2003).
\bibitem{LEP2}S. Schael et. al. (The ALEPH, DELPHI, L3, OPAL Collaborations and LEP Higgs Working Group),
 Eur. Phys. J. C 47, 547 (2006).
\bibitem{whi}M. Moretti, T. Ohl, and J. Reuter, Reports No. IKDA-2001-06 and No. LC-TOOL-2001-040, arXiv: hep-ph/0102195;
 W. Kilian, T. Ohl, and J. Reuter, Eur. Phys. J. C 71, 1742 (2011); \url{http://whizard.hepforge.org/}.
\bibitem{CIRCE2}W. Kilian et. al. (WHIZARD team), \url{http://whizard.hepforge.org/circe_files/CEPC/}.
\bibitem{BKG1}J. F. Gunion, T. Han, and R. Sobey, Phys. Lett. B 429, 79 (1998).
\bibitem{BKG2}X. Mo, G. Li, M.-Q. Ruan, and X.-C. Lou, Chin. Phys. C40, 033001 (2016);
 Z. Chen, Y. Yang, M. Ruan, D. Wang, G. Li, S. Jin, and Y. Ban, arXiv: 1601.05352.
\bibitem{BKG3}H. Li, arXiv: 1007.2999.
\bibitem{ptb}H. Li, PhD Thesis (Universit$\acute{\textrm{e}}$ de Paris-Sud, 2009), \url{http://hal.inria.fr/file/index/docid/430432/filename/Li.pdf}.
\bibitem{Br}The LHC Higgs Cross Section Working Group, arXiv: 1610.07922.
\bibitem{talk}Y.-N. Mao (2016), \url{http://indico.ihep.ac.cn/event/5600/session/114/contribution/428/material/slides/0.pdf}.
\bibitem{LHCrare}D. Curtin et. al., Phys. Rev. D 90, 075004 (2014).
\bibitem{EWP}M. Baak, J. Cuth, J. Haller, A. Hoecker, R. Kogler, K. M$\ddot{\textrm{o}}$nig, M. Schott, and J. Stelzer, Eur. Phys. J. C 74, 3046 (2014).


\end{thebibliography}
\end{document}